\documentclass[%
 aip,
 jmp,%
 amsmath,amssymb,
reprint,%
]{revtex4-1}

\usepackage{graphicx}
\usepackage{dcolumn}
\usepackage{bm}
\usepackage{subfigure}
\usepackage{siunitx}

\begin{document}


\title[Low RF power plasma ignition and detection for in-situ cleaning of 1.3 GHz 9-cell cavities]{Low RF power plasma ignition and detection for in-situ cleaning of 1.3 GHz 9-cell cavities}
\thanks{Work supported by D.O.E. Contract No. DE-AC02-07CH11359}

\author{P. Berrutti}
\email{berrutti@fnal.gov.}
\author{B. Giaccone}%
\author{M. Martinello}
\author{A. Grassellino}
\author{T. Khabiboulline}
\affiliation{ 
Fermi National Accelerator Laboratory, Batavia, Illinois 60510, USA
}%

\author{M. Doleans}
\author{S. Kim}
\affiliation{Oak Ridge National Laboratory, Oak Ridge, Tennessee 37831, USA
}

\author{D. Gonnella}
\author{G. Lanza}
\author{M. Ross}
\affiliation{SLAC National Accelerator Laboratory, Menlo Park, California 94025, USA
}%

\date{\today}

\begin{abstract}
Superconducting Radio Frequency (SRF) cavities performance preservation is crucial, from vertical test to accelerator operation. Field Emission (FE) is still one of the performance limiting factors to overcome and plasma cleaning has been proven successful by the Spallation Neutron Source (SNS), in cleaning field emitters and increasing the work function of Nb. A collaboration has been established between Fermi National Accelerator Laboratory (FNAL), SLAC National Accelerator Laboratory and Oak Ridge National Laboratory (ORNL) with the purpose of applying plasma processing to the Linac Coherent Light Source-II (LCLS-II) cavities, in order to minimize and overcome field emission without affecting the high Q of nitrogen-doped cavities. The cleaning for LCLS-II will follow the same plasma composition adopted at SNS, which allows in-situ processing of cavities installed in cryomodules from hydrocarbon contaminants. A novel method for plasma ignition has been developed at FNAL: a plasma glow discharge is ignited using high order modes to overcome limitations imposed by the fundamental power coupler, allowing in-situ cleaning for cavities in cryomodule. The plasma can be easily ignited and tuned in each of the cavity cells using low RF power. A method for plasma detection has been developed as well, which allows the detection of the plasma location in the cavity without the need of cameras at both cavity ends. The presented method can be applied to other multi-cell cavity designs, even for accelerators where the coupling for the fundamental modes at room temperature is very weak.

\end{abstract}

\maketitle

\section{\label{sec0:level1}Introduction}
Plasma processing allows in-situ cleaning of SRF cavities for field emission \cite{padamsee} mitigation purposes. The technique has been proven to be effective on SNS high beta (HB) cryomodules \cite{Dol_clean}\textsuperscript{,} \cite{Dol_RF}.
 A collaboration between ORNL, SLAC National Accelerator Laboratory  and Fermi National Accelerator Laboratory (FNAL) is working on adapting the plasma cleaning to the Linac Coherent Light Source-II (LCLS-II) cavities. The main objective is to reduce field emission (FE) without affecting the high $Q_{0}$ of the nitrogen-doped cavities \cite{grassellino2013nitrogen} implemented in LCLS-II cryomodules. 

In order to successfully perform in-situ cleaning of SRF cavities it is necessary to ignite the plasma, control its intensity and locate which cell is hosting the glow discharge, without having the possibility of looking inside the cavity RF volume. This article presents the complete procedure developed for in-situ cleaning of 1.3 GHz 9-cell TESLA-type cavities: plasma ignition, RF plasma detection and plasma tuning have been successfully implemented for the LCLS-II project. Two different low RF power plasma ignition methods are described, in both cases High Order Modes (HOMs) are exploited to ignite a glow discharge in the cavity. Both methods allow plasma ignition with just few Watts of RF power, since the cavity HOMs are strongly couple to the dampers also at room temperature.
Experimental results of both ignition methods are presented for Neon and Argon gas. In addition a study of electric field and RF voltage required to start a glow discharge is discussed. Experimental measurements of ignition curves of Neon and Argon are shown allowing for the optimization of parameters as forward power, pressure and gas. Plasma tuning measurements are presented in the text as a means to control plasma density. A method for plasma detection has been developed for 1.3 GHz cavities: it is based on Slater's theorem\cite{Slater} and allows monitoring the plasma location by measuring the RF frequencies of the first HOMs pass-band.

\section{\label{sec1:level1}Plasma Ignition studies}
The plasma processing technique requires the ignition of a glow discharge within the cavity RF volume. The cavity is filled with about 200\,mTorr of Neon and a small percentage of Oxygen is introduced in the cavity while the plasma is ignited. The plasma is kept glowing until the exhaust gas shows negligible traces of hydrocarbons byproducts, this process is repeated for all the cavity cells. 

\begin{figure}
   \centering
   \includegraphics[scale=0.6]{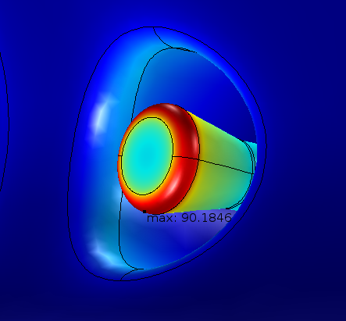}
   \caption{LCLS-II coupler $E_{peak}$ in kV/m for cavity surface field of 10 kV/m.}
   \label{peak_field_coupler}
\end{figure}

The studies of plasma ignition started from ORNL expertise on high beta (HB) SNS cavities, the adaptation of the dual tone excitation to LCLS-II was a natural consequence of ORNL successful prior experience. The dual tone excitation \cite{Dol_RF} consists in superimposing two modes from the fundamental pass-band in order to generate asymmetry in the field distribution between cavity cells. The plasma will then glow in the cell where the ignition level has been reached. This technique works extremely well for SNS cavities, it has been proven reliable and several cryomodules have been processed at ORNL. At room temperature SNS high beta cavities have a $Q_{0}\approx 10^4$ while the power coupler has $Q_{ext}\approx 7 \times 10^5$ for the operating mode\cite{SNS_Qext}. LCLS-II cavities have approximately the same $Q_{0}$, at room temperature, while the $Q_{ext} \approx 4 \times 10^7$ of the power coupler\cite{LCLSII_FDR} is set to roughly 60 times the SNS value. Lower coupling implies more forward power needed to ignite the plasma discharge and higher peak field on the coupler tip, hence higher risk of plasma ignition in the antenna. The ratio between peak field in LCLS-II coupler and in the cavity, suggested that the risk of antenna ignition is very high. For a peak field of 10\,kV/m in cavity, the peak value at the coupler is 90\,kV/m, see Fig.~\ref{peak_field_coupler}. Since plasma ignition in the coupler has several negative effects, from arcing to deposition of metal particles on the ceramic window, an alternative method has been developed at FNAL. \\
The adaptation of the in-situ plasma cleaning to HWR cavities is being studied at Institute of Modern Physics, CAS\cite{Wu_plasma}. The assembly of the HWR cavity for plasma ignition requires a special variable coupler in order to reduce the mismatch between the $Q_{ext}$ of antenna and the $Q_{0}$ of the cavity at room temperature. This highlights the uniqueness of the new procedure developed at FNAL that allows plasma ignition with HOMs using only a few watts of forwarded power.

\subsection{\label{sec2:level2} HOMs Plasma Ignition}
\begin{figure}
   \centering
 \subfigure[]
   {\includegraphics[scale=0.27]{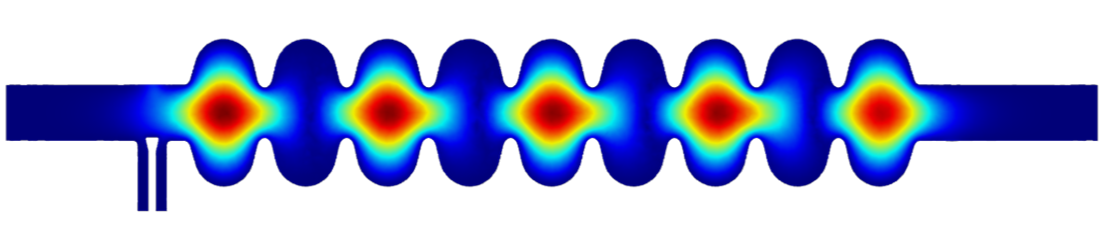}\label{E_dist_sustain_1}}
 \subfigure[]
   {\includegraphics[scale=0.275]{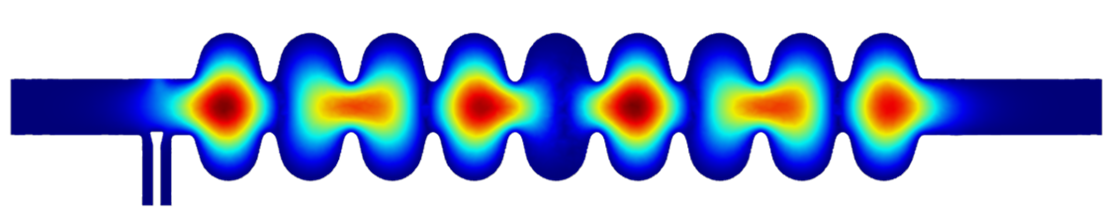}\label{E_dist_sustain_2}}
 \subfigure[]
   {\includegraphics[scale=0.27]{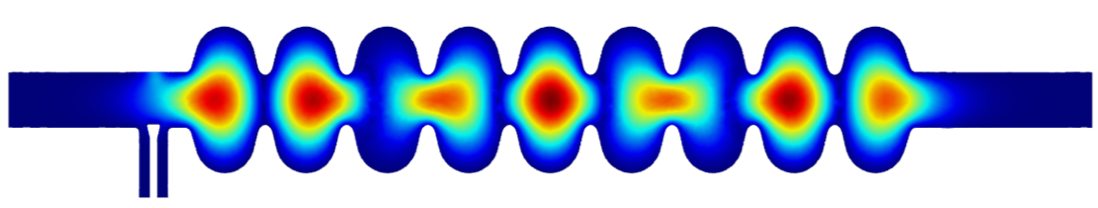}\label{E_dist_sustain_3}}
 \caption{Electric field distribution on X-Z plane of set of modes used to sustain the glow discharge in all cells: 1-5 for all odd cells (a), 1-6 in cells 4 and 6 (b) and 1-7 for cells 2 and 8 (c).}
   \label{E_dist_sustain}
\end{figure}

 The alternative studies for plasma ignition involve HOMs of 1.3 GHz cavities \cite{Berr_RF}, several modes and pass-bands have been analyzed and dipole modes have been identified as good candidates for plasma ignition. The HOMs plasma ignition technique is very efficient and requires only few Watts of RF power, overcoming limitations imposed by the poor coupling of the fundamental pass-band. First and second dipole pass-bands are strongly coupled through both HOM couplers at room temperature: reflection coefficient $|\Gamma|^2\approx0.01-0.3$ while for the fundamental pass band $|\Gamma|^2\approx0.99$. The first two dipole bands are hybrid in Tesla cavities: for some modes the field distribution is TE-like in some cells and TM-like in others. Given the complexity of the field distribution and the impossibility of modeling analytically these modes like the $\mathrm{TM_{010}}$ pass-band, finite element simulations are needed. Some of the dipole modes have a field distribution similar to monopoles and therefore they are suitable for igniting plasma in the whole cell. For example, see Fig.~\ref{E_dist_sustain_1} the $\mathrm{5^{th}}$ mode from the $\mathrm{1^{st}}$ dipole pass-band is suitable to sustain plasma in all odd cells.
 The field profiles of the cavity's eigenmodes are simulated with COMSOL Multiphysics\textsuperscript{\textregistered} \cite{COMSOL} using the frequency domain technique.
 
 The first method is HOMs superposition, which is equivalent to ORNL dual tone excitation for SNS HB cavities \cite{Dol_RF}. The second technique, called plasma bridging, allows transferring the plasma between neighboring cells, after ignition of a known cavity cell. This has the advantage of never requiring the plasma to be shut off during the cleaning process, ensuring no interruptions in the contaminants removal. 
 
HOMs superposition is the first plasma ignition method developed for 1.3 GHz cavities and it consists in the superposition of two modes: one to build up field and another to generate asymmetry making plasma ignition more favorable only in one cell. Usually modes from the $\mathrm{2^{nd}}$ dipole pass-band can build up electric field in a couple of cells while the $\mathrm{1^{st}}$ dipole pass-band is used to give asymmetry to the total field distribution. For sake of simplicity one can refer to modes of the $\mathrm{1^{st}}$ dipole pass-band as 1-X and modes from the $\mathrm{2^{nd}}$ dipole pass-band as 2-Y where X and Y are mode numbers from 1 to 9. Using the electric field profile along z from simulations it is possible to find linear combinations of modes (1-X and 2-Y) that will ignite a selected cell in the cavity.

\begin{figure}
      \centering
 \subfigure[]
   {\includegraphics[scale=0.3]{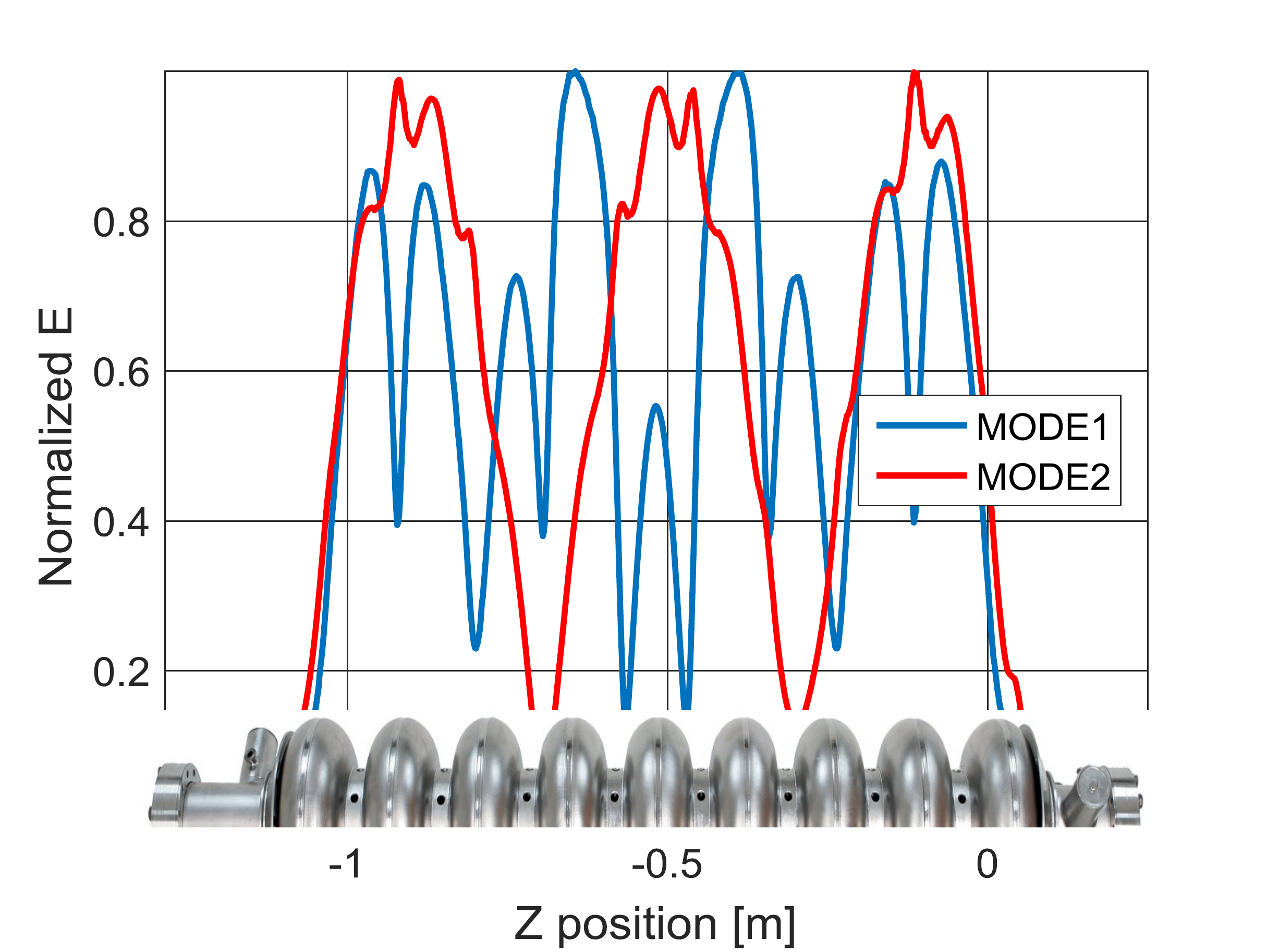}}
 \subfigure[]
   {\includegraphics[scale=0.3]{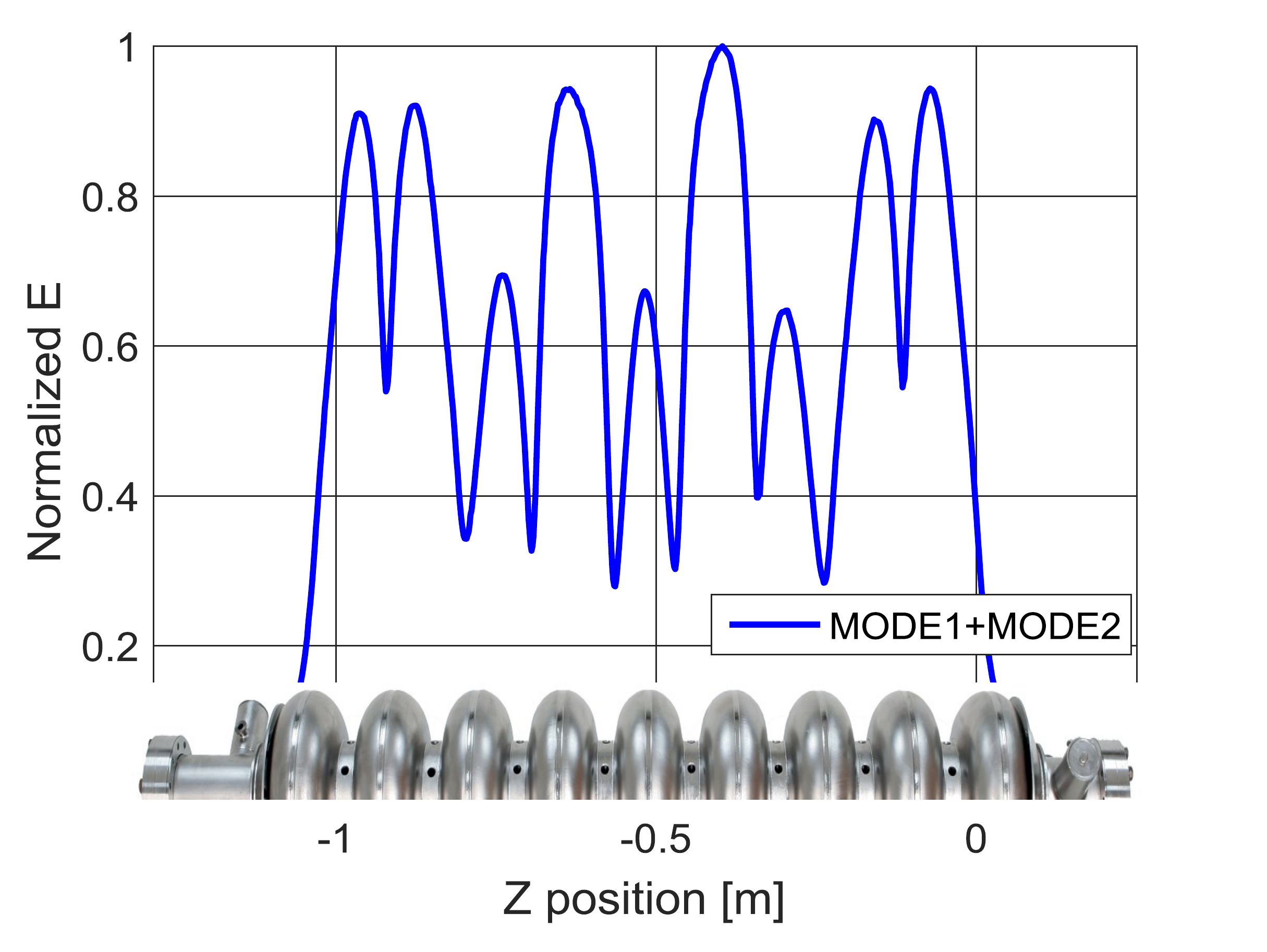}}
 \caption{Electric field distribution of HOMs used to ignite cell 6 (a) and final field distribution after superposition (b).}
   \label{HOMS_fields_addition}
\end{figure}

An example of selective cell ignition is reported in Fig.~\ref{HOMS_fields_addition}: the field distribution of MODE1 in Fig.~\ref{HOMS_fields_addition}(a) would ignite either cell 4 or 6, one can generate asymmetry by superposing MODE2 in the same figure which creates asymmetry between cell 4 and 6.
As a result of mode superposition the highest electric field is located in cell 6, shown in  Fig.~\ref{HOMS_fields_addition}, and therefore a plasma discharge is going to ignite preferably at this location. Using the field distribution from RF simulations it is possible to find which combination of HOMs ignites each cell in LCLS-II cavities. For all cells, exception made for cell 5,  two HOMs are needed for plasma ignition and the proper amplitudes can be calculated using field distribution from RF simulations. The HOMs superposition for plasma ignition has been proven effective and results are presented in section \ref{selective_cell_ignition}.

\begin{figure}
   \centering
   \includegraphics[scale=0.27]{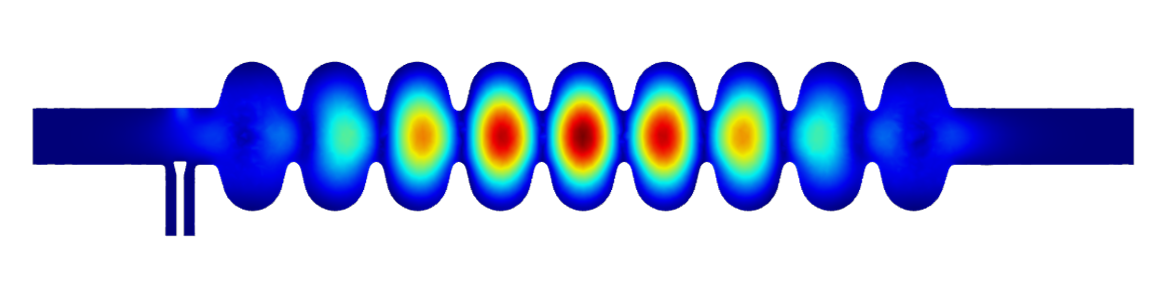}
   \caption{Electric field distribution on X-Z plane of the $\mathrm{1^{st}}$ mode from the $\mathrm{2^{nd}}$ dipole pass-band, used for ignition of cell 5.}
   \label{2-1_field_XZ}
\end{figure}

\begin{figure}
   \centering
 \subfigure[]
   {\includegraphics[scale=0.275]{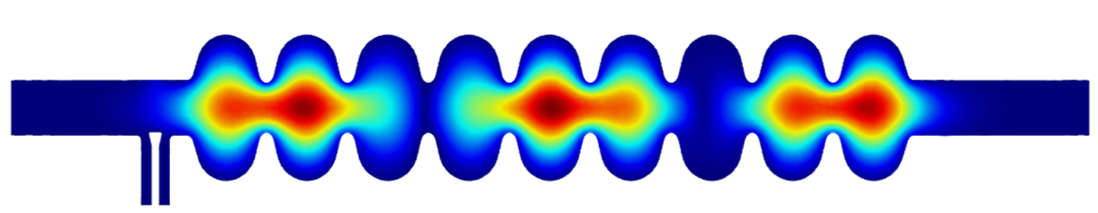}\label{transfer_modes_E_1}}
 \subfigure[]
   {\includegraphics[scale=0.27]{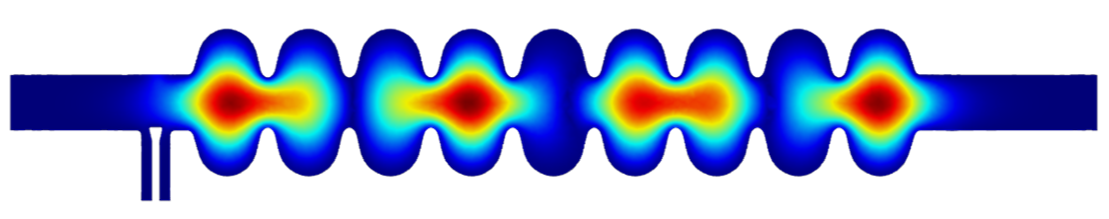}\label{transfer_modes_E_2}}
 \subfigure[]
   {\includegraphics[scale=0.27]{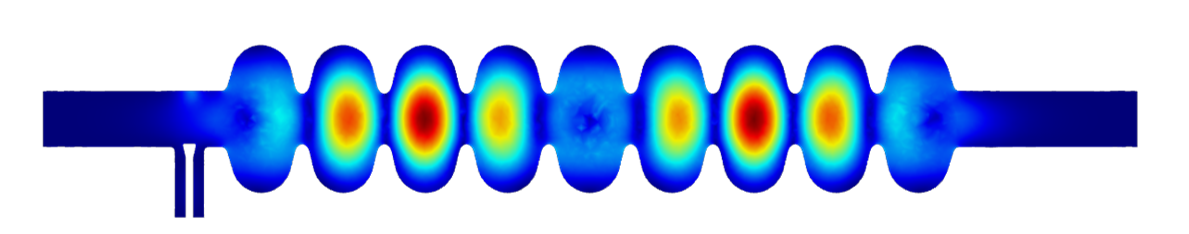}\label{transfer_modes_E_3}}
 \caption{Electric field distribution on X-Z plane of all modes used for plasma transfer between cells: 1-3 (a), 1-4 (b) and 2-2 (c).}
   \label{transfer_modes_E}
\end{figure}

The second plasma ignition technique, called plasma bridging,consists in igniting a known cell in the cavity, then transferring the glow discharge to all other cells, one at the time. This technique takes advantage of the broad field distribution of some HOMs which allow igniting volume shared by neighboring cells making the bridging of the plasma very easy and reliable. Mode 2-1 has been chosen for the initial ignition in order to make the ignition method robust and dependable: only cell 5 will be ignited as one can see from Fig.~\ref{2-1_field_XZ}. Three modes are needed to transfer the plasma to all other cells (modes 1-3, 1-4 and 2-2 shown in Fig.~\ref{transfer_modes_E}). Additional 3 modes are used to sustain and tune the plasma in each cell: 1-5 for all odd cells Fig.~\ref{E_dist_sustain_1}, 1-6 for cells 4 and 6 Fig.~\ref{E_dist_sustain_2}, and 1-7 for cells 2 and 8 Fig.~\ref{E_dist_sustain_3}.

After ignition of cell 5 the plasma can be locked into the same cell superimposing mode 1-5 then shutting off mode 2-1. The plasma can be transferred to either cell 4 or 6 superimposing mode 1-3 or 1-4 respectively; then mode 1-5 can be shut off and 1-6 can be used to lock the plasma into either cell 4 or cell 6. One can transfer and stabilize the plasma from cell M to either neighboring cells by following the procedure described below:
\begin{itemize}
\item[1]{Stabilize plasma in cell M with MODE1.}
\item[2]{Superimpose MODE2 to bridge cell M and N}
\item[3]{Turn off MODE1}
\item[4]{Turn on MODE3 to lock plasma in cell N}
\end{itemize}

Table \ref{plasma_bridge_modes} summarizes all needed modes to bridge plasma between all cells: from cell 1 to 9 and from cell 9 to 1. Plasma bridging has been verified experimentally, the procedure is dependable and working on several 1.3 GHz cavities; results are presented in section \ref{selective_cell_ignition}.
\begin{table}
   \centering
   \caption{Modes used in plasma bridging for 1.3 GHz cavities: first plasma ignition in cell 5, plasma is transferred to all other cells afterwards.}
   \begin{ruledtabular}
   \begin{tabular}{|c|c|r|r|r|}
      \textbf{CELL M}&\textbf{CELL N}&\textbf{MODE1}&\textbf{MODE2}&\textbf{MODE3}\\
\hline
      \textbf{5}& \textbf{5} & 2-1  & - &1-5\\
      \textbf{5}& \textbf{4} & 1-5  & 1-4 &1-6\\  
      \textbf{4}& \textbf{3} & 1-6  & 1-3 &1-5\\
      \textbf{3}& \textbf{2} & 1-5  & 2-2 &1-7\\
      \textbf{2}& \textbf{1} & 1-7  & 1-3 &1-5\\
      \textbf{1}& \textbf{2} & 1-5  & 1-3 &1-7\\
      \textbf{2}& \textbf{3} & 1-7  & 2-2 &1-5\\
      \textbf{3}& \textbf{4} & 1-5  & 1-4 &1-6\\
      \textbf{4}& \textbf{5} & 1-6  & 1-3 &1-5\\
      \textbf{5}& \textbf{6} & 1-5  & 1-3 &1-6\\
      \textbf{6}& \textbf{7} & 1-6  & 1-4 &1-5\\
      \textbf{7}& \textbf{8} & 1-5  & 2-2 &1-7\\
      \textbf{8}& \textbf{9} & 1-7  & 1-3 &1-5\\
      \textbf{9}& \textbf{8} & 1-5  & 1-3 &1-7\\
      \textbf{8}& \textbf{7} & 1-7  & 2-2 &1-5\\
      \textbf{7}& \textbf{6} & 1-5  & 1-4 &1-6\\
      \textbf{6}& \textbf{5} & 1-6  & 1-3 &1-5\\
   \end{tabular}
   \end{ruledtabular}
   \label{plasma_bridge_modes}
\end{table}

\begin{table*}
   \centering
   \caption{HOMs plasma ignition for 9-cell LCLS-II cavity at 200 mTorr Ar: modes and power in Watts needed to ignite each cell.}
   \begin{ruledtabular}
   \begin{tabular}{|l|c|c|c|c|c|}
       \textbf{CELL} &\textbf{MODE1} & \textbf{MODE2}                      & $\mathbf{P_{Fwd1}}$ \textbf{Calc}& \ $\mathbf{P_{Fwd1}}$ \textbf{Meas}& $\mathbf{P_{Tot}}$ \\
\hline
           \textbf{1}         & 2-4 & 1-6  & 1.50 &1.49 & 4.71     \\
           \textbf{2}       & 2-6  & 1-4 & 5.46 &5.45 &8.97 \\ 
           \textbf{3}        & 2-2 & 1-3& 4.85 &4.90 &6.35 \\ 
           \textbf{4}       & 2-5 & 1-4&1.13 &1.13 &5.89 \\
           \textbf{5}     &  2-1  &  & 2.97 &2.97 &2.97 \\
           \textbf{6}       & 2-5 & 1-3 & 3.90 &3.91 &7.78 \\
           \textbf{7}       & 2-2 & 1-4& 3.87 &3.85 &6.02 \\
           \textbf{8}       & 2-6 & 1-9& 3.98 &4.00 &7.23 \\
           \textbf{9}       & 2-4 & 1-4& 1.48 &3.86 &7.28 \\
   \end{tabular}
   \end{ruledtabular}
   \label{table_pw_ign}
\end{table*}

\section{\label{sec3:level1}Experimental setup}
\begin{figure*}
   \centering
   \includegraphics[scale=0.7]{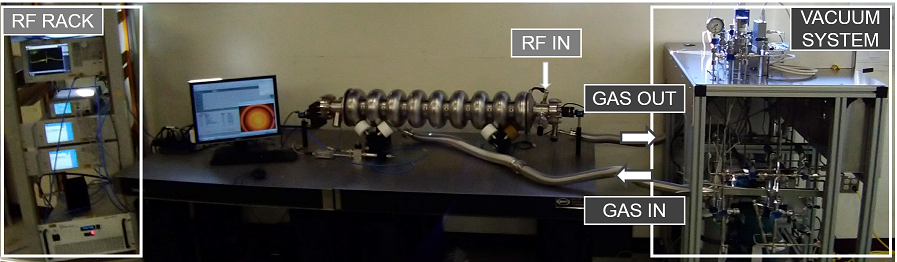}
   \caption{Plasma cleaning experimental RF and vacuum systems connected to a 9-cell LCLS-II cavity.}
   \label{Plasma_systems}
\end{figure*}

The plasma cleaning RF setup consists in: vector network analyzer (VNA), two signal generators, RF power amplifier, bidirectional coupler and power meters. The VNA is used to measure and track cavity frequency prior and during plasma ignition via S21 measurements through HOM couplers. Two signal generators are needed for mode superposition as they generate the input of the RF power amplifier. The RF signal goes to one of the cavity HOM couplers after going through a bi-directional coupler, which allows to take measurements of forward and reflected power. The setup is then completed with cameras on each side of the cavity to visually locate the plasma. The plasma processing vacuum system allows to control the gas flow through the cavity being processed: the supply of gases is controlled by needle valves or a mass flow controller,  the by-products exiting through the exhaust line are analyzed by a RGA. The experimental setup is shown in Fig.~\ref{Plasma_systems}: RF rack, RGA and cameras can be controlled by the operator on a computer, making plasma processing operations faster.
\section{\label{sec4:level1}Plasma Ignition Experiments}
The first plasma ignition experiments on LCLS-II cavities were carried out to prove feasibility of ignition using HOMs and HOM couplers. Prior to the experiments at FNAL plasma ignition has been tried only with fundamental pass-band mode of SRF multi-cell cavities \cite{Dol_RF}\textsuperscript{,} \cite{JLAB_plasma}\textsuperscript{,} \cite{Eng_1.3_plasma}. Ignition power and its dependence on the gas type and pressure have been studied and the results for LCLS-II cavities with Ne and Ar are presented in the following section. The power needed to ignite a plasma discharge is considerably low thanks to the good coupling of the HOMs selected for ignition. This should allow plasma ignition in cryomodule environment where the forward power is limited to approximately 10 Watts by cables which are low power. Another set of ignition experiments has been carried out to prove it is possible to control and select which cell would be hosting the plasma glow discharge: both HOMs superposition and Plasma bridging have given successful results, which are presented in section \ref{selective_cell_ignition}.

\subsection{Ignition curves}
Different pressures and gases have been used for plasma ignition\cite{brown1966introduction}, using modes from the first and second dipole pass-bands. These studies will help defining the recipe for plasma processing LCLS-II cavities.
Argon and Neon have been chosen for these tests because they have both been used at ORNL during the plasma processing development for SNS HB cryomodules. Ar requires lower power than Ne to ignite a plasma glow discharge since it has lower ionization energy (15.8\,eV for Ar, 21.6\,eV for Ne).\\ 
The studies started at 200\,mTorr for both gases, in order to reproduce the SNS procedure. Other pressures were used as well: higher pressures allow to decrease the amount of power needed to ignite the discharge but can lead to an unstable plasma which proves difficult to transfer from one cell to the other, or has a higher probability to ignite the HOM coupler; lower pressures show a progressive increase in the power of ignition. For Ne at 50\,mTorr it was not possible to ignite the plasma.
For these reasons pressures higher than 300\,mTorr for Ne (350\,mTorr for Ar) and lower than 50\,mTorr for Ne (30\,mTorr for Ar) were not object of study.

Each mode of the $1^{st}$ and $2^{nd}$ dipole pass-band is sent to the cavity and $\mathrm{P_{FWD}}$ is slowly increased until plasma glow discharge is ignited. $\mathrm{P_{FWD}}$, $\mathrm{P_{R}}$, $\mathrm{P_{T}}$ are measured just before the glow discharge, because the plasma changes the dielectric constant of the gas. This results in a frequency shift of the resonance peak and in a drop of the quality factor.

\begin{figure}
	\centering
	\includegraphics[scale=0.35]{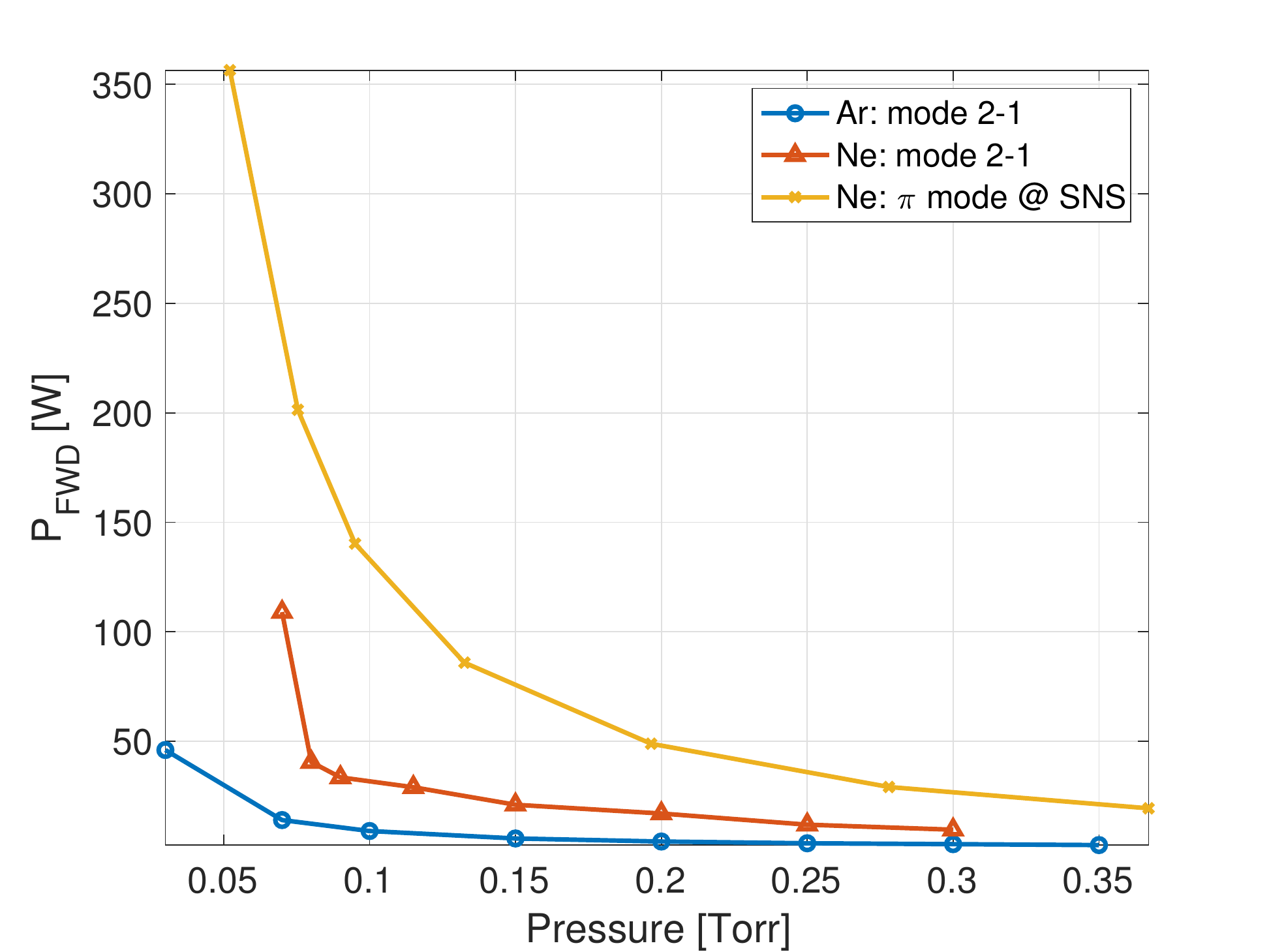}
	\caption{Curves of $\mathrm{P_{FWD}}$ necessary to ignite the plasma glow discharge for mode 2-1 as a function of the pressure for Argon and Neon. The plot also shows the curve of ignition of the $\pi$ mode  for SNS cavities.}
	\label{fig:21vsSNS}
\end{figure}

\begin{figure}
	\includegraphics[scale=0.35]{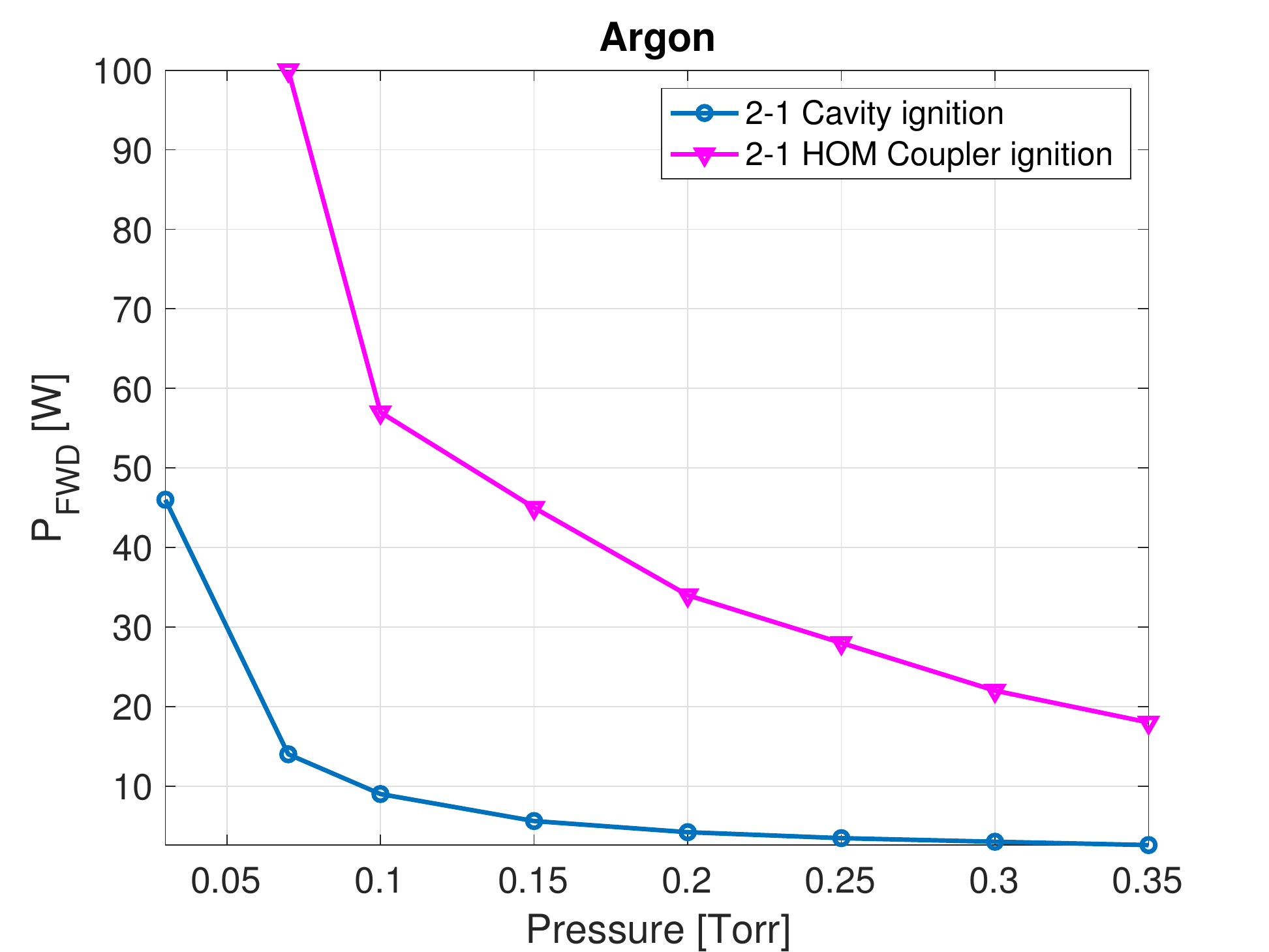}
	\caption{$\mathrm{P_{FWD}}$ required to ignite the plasma in the cavity and in the HOM coupler for Argon gas.}
	\label{fig:cav_vs_coupl}
\end{figure}

The possibility of transferring the plasma from one cell to the other and to tune its intensity has been regularly checked.
When possible, the ignition curve of plasma in the HOM coupler has been measured. This curve is particularly relevant for Ar, since its low power of ignition could lead to RF instability, increasing the probability of accidentally igniting the antenna during the plasma cleaning procedure. 
Fig.~\ref{fig:21vsSNS} shows the curve of ignition, comparing Ar and Ne for the first mode of the $2^{nd}$ dipole pass-band.
The comparison between the curve of ignition of Neon for mode 2-1 and the $\pi$ mode of SNS cavities highlights how the use of HOMs is more efficient. This method allows to ignite the plasma glow discharge with only a few watts, thanks to the favorable coupling between the cavity and the HOM ports at room temperature. 
In addition, Fig.~\ref{fig:cav_vs_coupl} shows that the risk of igniting the plasma in the HOM coupler is negligible, since its power of ignition is at least six times higher that the ignition in the cavity.


It is possible to extract the values of voltage and field gradient, that correspond to the plasma glow discharge, from the measured values of $\mathrm{P_{FWD}}$, $\mathrm{P_{R}}$, $\mathrm{P_{T}}$.

\begin{figure}
	\centering
	\subfigure[\label{fig:field_21}]
	{\includegraphics[width=150pt]{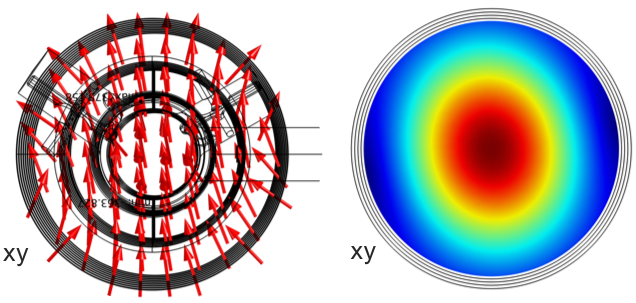}}\quad
	\subfigure[\label{fig:field_23}]
	{\includegraphics[width=150pt]{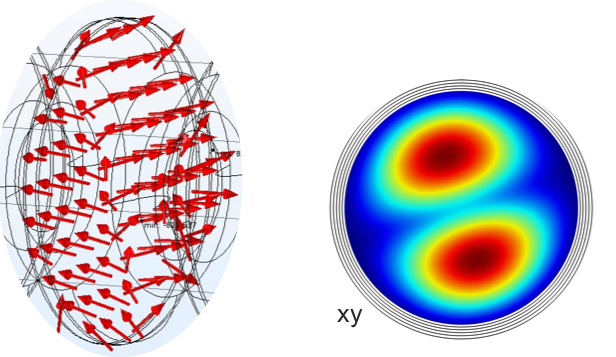}} \quad
	\caption{Field profiles simulated with COMSOL Multiphysics\textsuperscript{\textregistered}. Figure~\ref{fig:field_21} shows the field profile for mode 2-1 (belonging to the $1^{st}$ group): when forwarded to the cavity it ignites plasma in cell $\#$5; its electric field is directed on the $y$ axis. Figure~\ref{fig:field_23} shows the field profile for mode 2-3 (belonging to the $2^{nd}$ group): it ignites cell $\#$6 where the E field is directed parallel to the $z$ axis.}
	\label{fig:COMSOLfield}
\end{figure}

Field profiles of the resonant modes are extracted from finite element simulations to calculate the simulated $\mathrm{P_{C}}$, necessary to obtain $\mathrm{V}$ and field gradient. Depending on the direction of the electric field in the cell of ignition, dipoles modes can be divided in two groups:
\begin{itemize}
    \item $z$ modes: maximum field directed parallel to the longitudinal axis ($z$ axis) of the cavity, modes 2-3 to 2-6 belong to this group and their accelerating gap is $\approx$ 9\,cm;
    \item $y$ modes: maximum field in the plane orthogonal to the longitudinal axis of the cavity, modes 1-2 to 1-8, 2-1, 2-2 belong to this second category and their accelerating gap is $\approx$ 20\,cm.
\end{itemize}
Figure~\ref{fig:COMSOLfield} shows an example of field profile for a mode belonging to the first group (mode 2-1) and to the second (mode 2-3). 

\begin{figure*}
	\centering
		\subfigure[\label{fig:Acc_grad_Ne}]
	{\includegraphics[width=150pt]{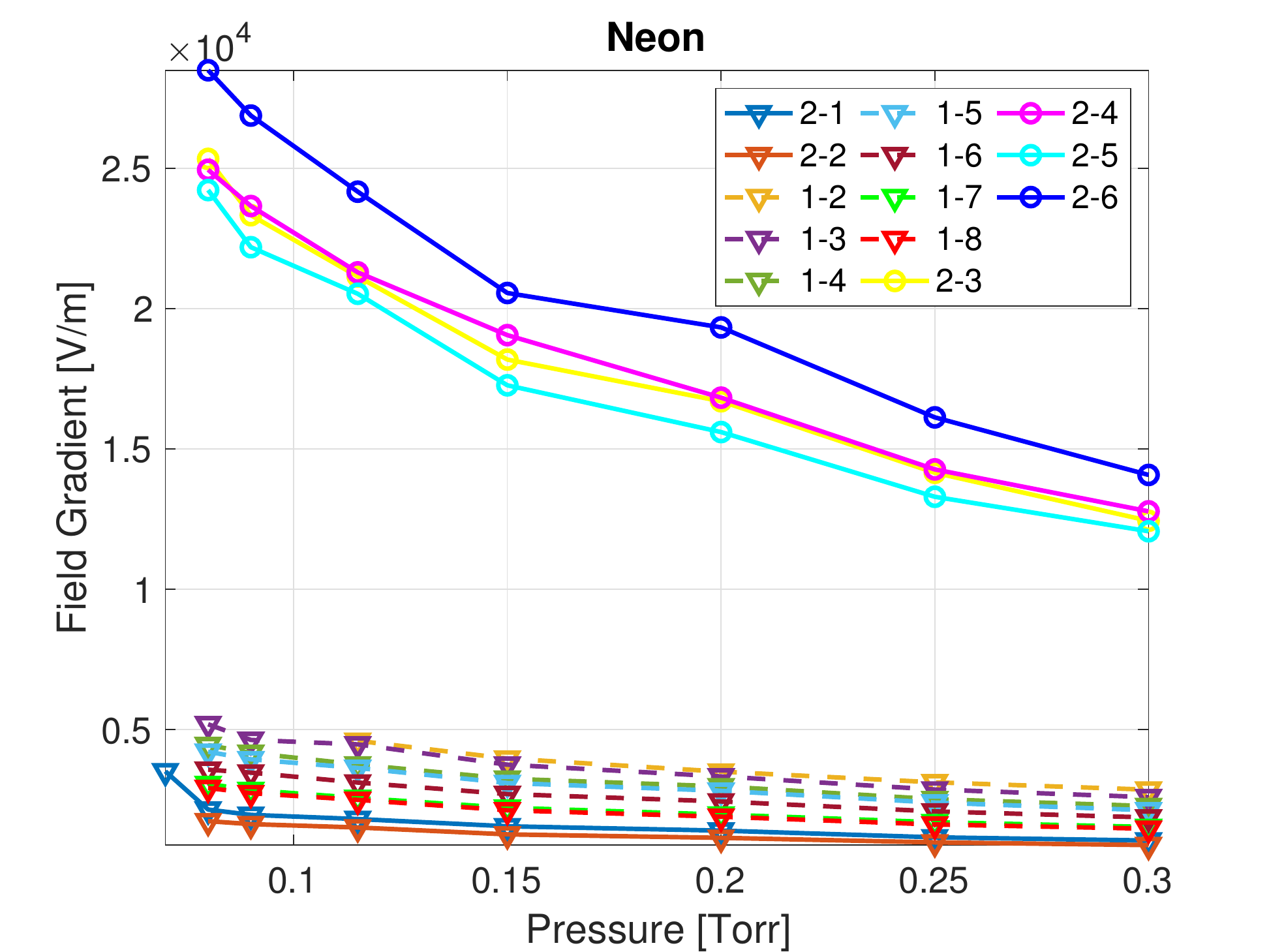}}
			\subfigure[\label{fig:Acc_grad_Ar}]
	{\includegraphics[width=150pt]{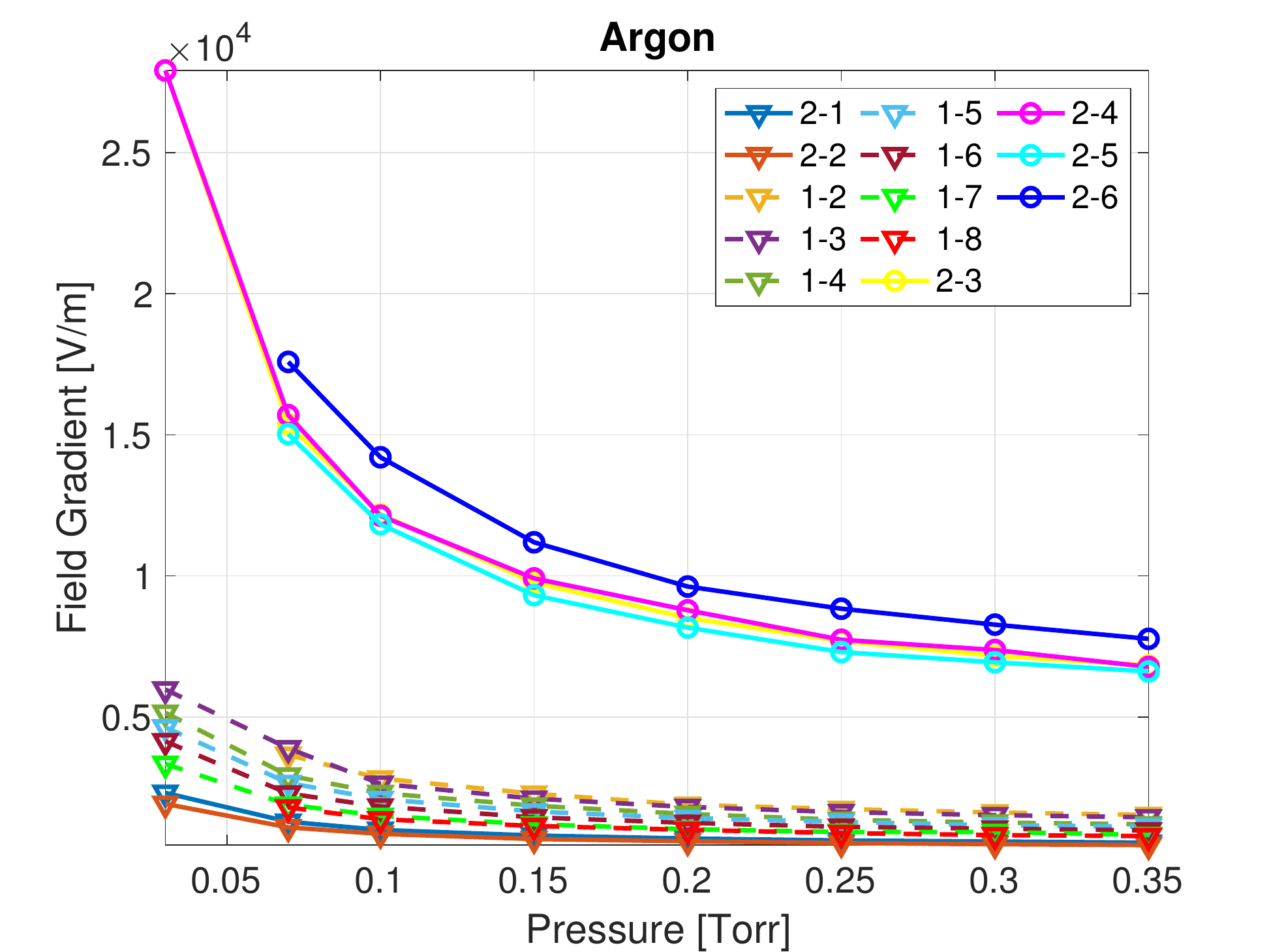}}\quad
		\subfigure[\label{fig:Voltage_Ne}]
	{\includegraphics[width=150pt]{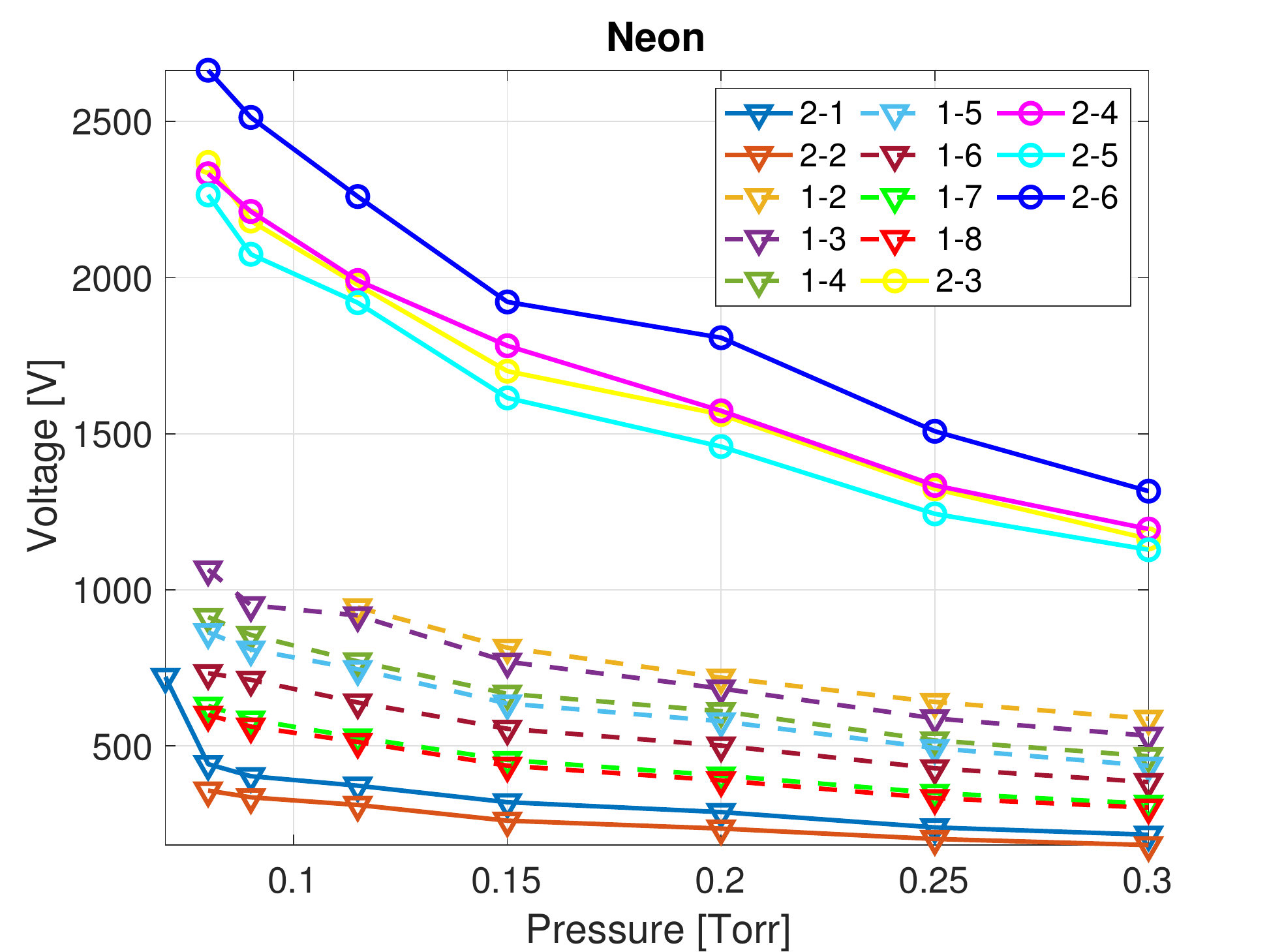}}
		\subfigure[\label{fig:Voltage_Ar}]
	{\includegraphics[width=150pt]{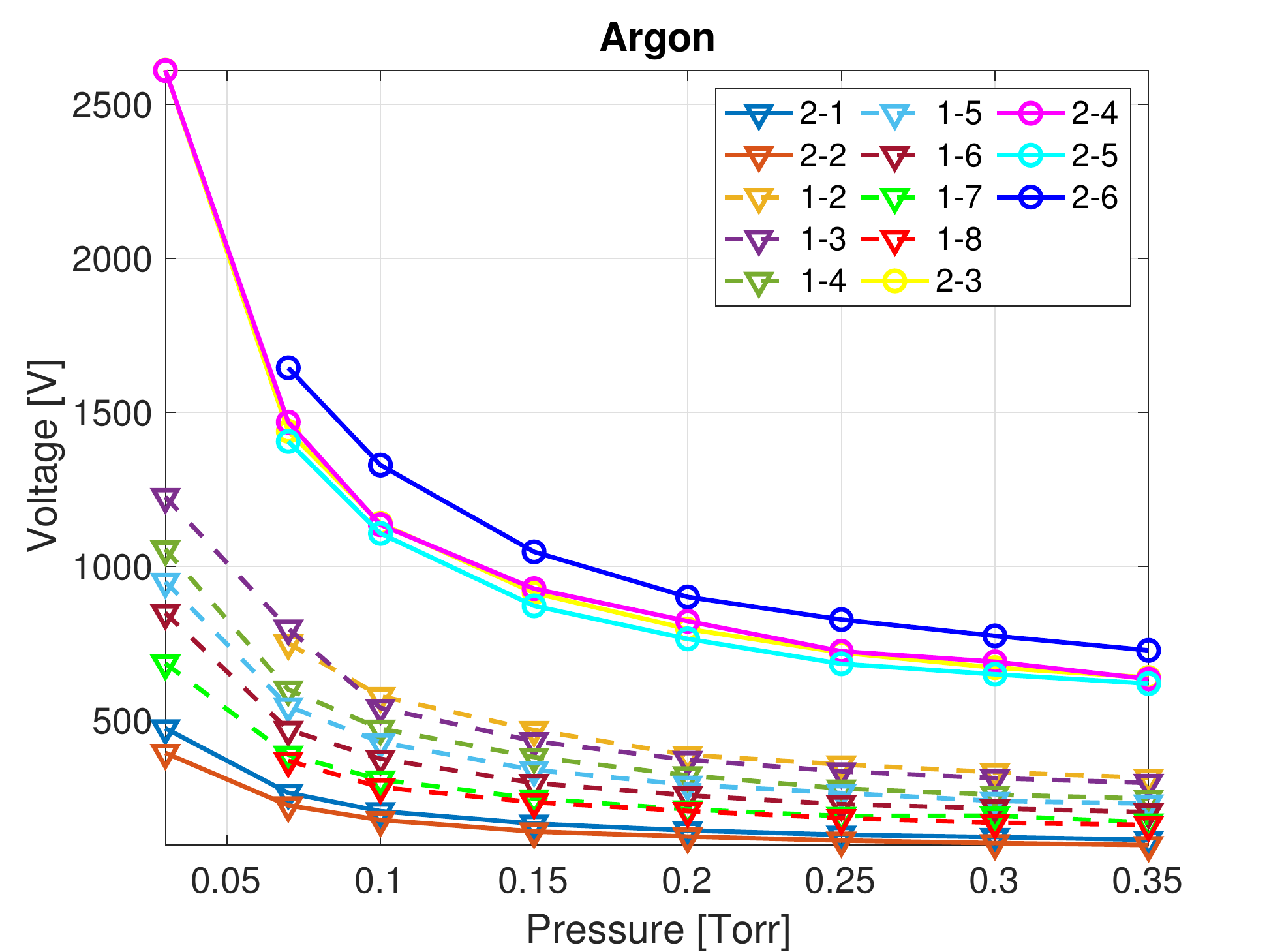}}\quad
	\caption{In figures \ref{fig:Acc_grad_Ne} and \ref{fig:Acc_grad_Ar} are plotted the curves of field gradient versus pressure for each mode of the first and second dipole pass-bands, for Ne and Ar respectively. Figures \ref{fig:Voltage_Ne} and \ref{fig:Voltage_Ar} show the curves of voltage versus pressure.}
	\label{fig:results_Ar_Ne}
\end{figure*}

Using the length of the accelerating gap, the maximum component of simulated field profile it is possible to calculate the simulated voltage. The experimental voltage is obtained from the simulated one using the measured $\mathrm{P_{C}}$.
The field gradient can be extracted simply dividing the calculated voltage for the length of the accelerating gap.
The resulting curves are shown in Fig.~\ref{fig:results_Ar_Ne} for Neon and Argon. Fig.~\ref{fig:Acc_grad_Ne} and \ref{fig:Acc_grad_Ar} show the field gradient as a function of the pressure; it can be seen that the curves are divided in two groups depending on the length of the gap: modes with shorter gaps present higher gradient field.
Fig.~\ref{fig:Voltage_Ne} and \ref{fig:Voltage_Ar} instead show the curves of voltage against pressure: the separation between the two groups of curves is reduced but still present. This effect could depend on the difference in frequency between the resonant modes.

The electric field profile of the $\pi$ mode has been exported from COMSOL Multiphysics\textsuperscript{\textregistered} simulations of the cavity. Following the same procedure, the simulated voltage is obtained using the length of one cell ($\approx$ 11\,cm) as the accelerating gap. Modes directed on the $z$ axis (2-3 to 2-6) have been used to estimate the power dissipated in the cavity and the experimental voltage at the moment of ignition.

\begin{figure}
	\centering
	{\includegraphics[scale=0.4]{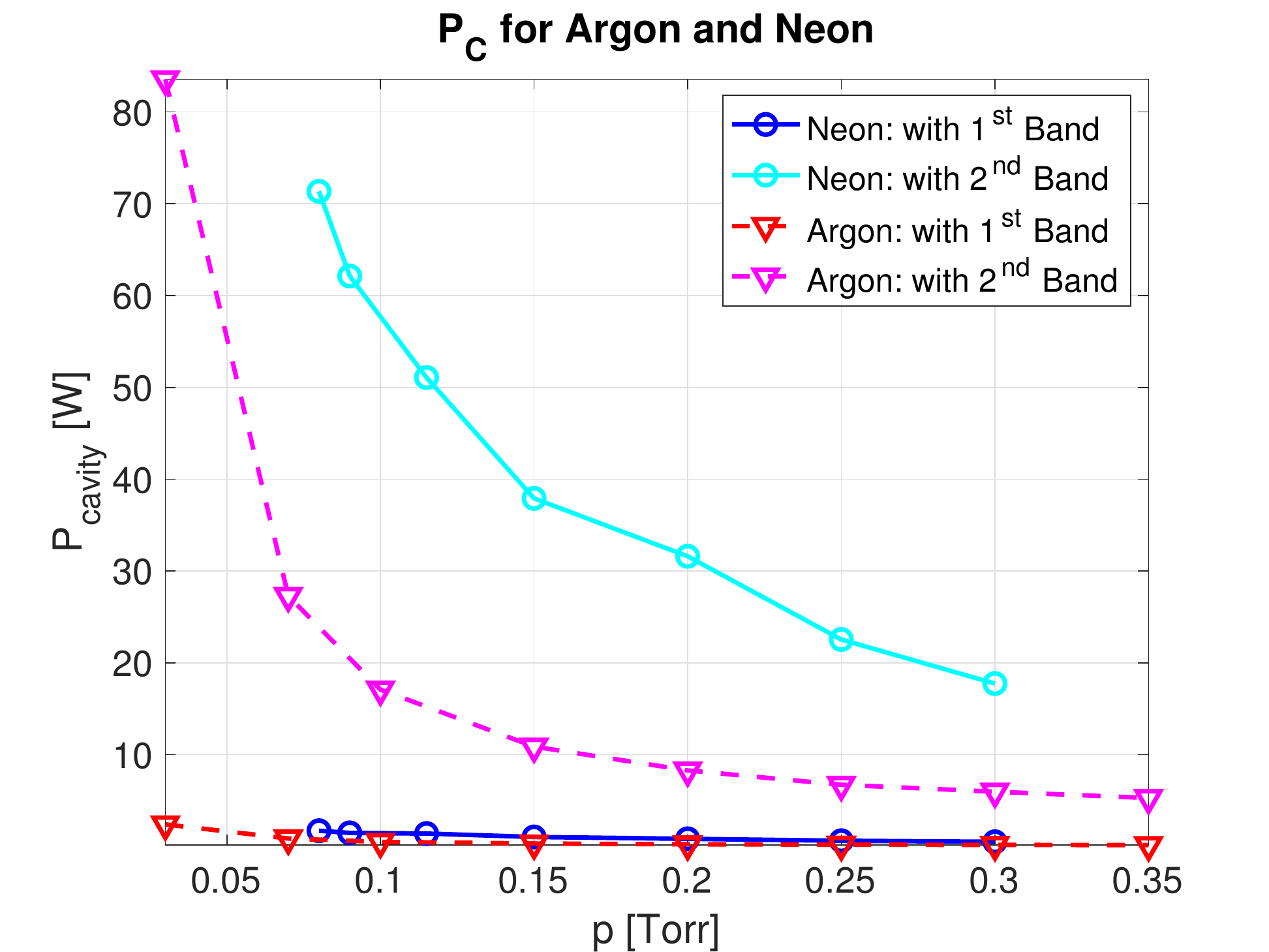}}
	\caption{Plot of the estimated $\mathrm{P_C}$ necessary for the $\pi$ mode to ignite the plasma in one cell. For each gas, the two curves show the maximum and minimum limit estimated using second and first dipole pass-bands, respectively.}
	\label{fig:pimode}
\end{figure}

The shortcoming of this method is the difference in frequency between the $\pi$ mode ($\SI{1.3}{\giga\hertz}$) and modes 2-3 to 2-6 (with frequencies $\approx \SI{1.85}{\giga\hertz}$). Modes of the $1^{st}$ dipole pass-band (1-2 to 1-8) have been used to have a second estimate of the $\mathrm{P_{C}}$ and the field gradient of the $\pi$ mode. The first dipole pass-band is closer in frequency to $\SI{1.3}{\giga\hertz}$, but the accelerating gap is almost double than $\pi$'s, since these modes are directed on the $y$ axis. The results of these two estimate are shown in Fig.~\ref{fig:pimode}, they can be interpreted as the maximum ($2^{nd}$ pass-band) and minimum ($1^{st}$ pass-band) limit for the $\mathrm{P_C}$ corresponding to plasma ignition using the $\pi$ mode.

Taking into account the low coupling at room temperature with the FPC (reflection coefficient $|\Gamma|^2\approx0.99$), the amount of power to be forwarded to the cavity to ignite plasma with the fundamental pass-band would be considerably high, requiring from hundreds to thousands of watts. In addition, the probability of igniting the glow discharge in the coupler increases because of field enhancement at the coupler, as mentioned in \ref{sec1:level1}.
This underlines the importance of using HOMs, and dipole pass-bands, to ignite the plasma glow discharge in LCLS-II cavities, instead of the fundamental pass-band.

\subsection{\label{selective_cell_ignition}Selective cell ignition experiments} 
\begin{figure}
   \centering
   \includegraphics*[scale=0.8]{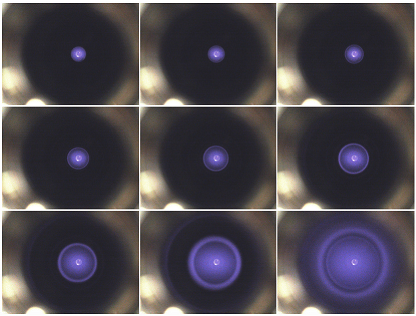}
   \caption{Ar plasma in each cells (9 to 1) of a LCLS-II cavity, fundamental power coupler side view.}
   \label{cells_ignited}
   \includegraphics*[scale=0.8]{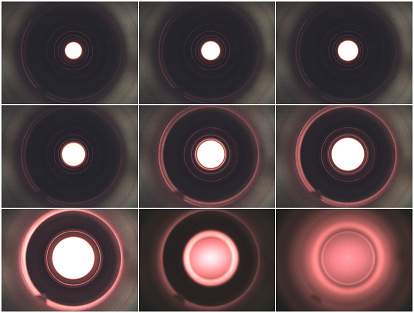}
   \caption{Ne plasma in each cells (9 to 1) of a LCLS-II cavity, FPC side view.}
   \label{Ne_cell_1_9}
\end{figure}

The first successful HOMs superposition plasma ignition experiment on 1.3 GHz 9-cell cavities has been carried out at FNAL. Thanks to the extremely favorable coupling of the HOM couplers for LCLS-II project only few watts are needed to ignited a glow discharge in the cavity. Argon has been used, in place of Neon, at a pressure of 200 mTorr, which matches the working pressure for SNS cavities plasma cleaning. Fig.~\ref{cells_ignited} shows Ar plasma in LCLS-II cavity from cell 1 (top left) to 9 (bottom right).

Only modes from the first and second dipole pass-bands have been used, and for all the modes used from the second pass-band it was possible to identify exactly the ignition power. Using the field distribution from simulation it is possible to calculate amplitude, coefficients for all modes needed. Combining these two one can calculate exactly how much power is needed from the second pass-band. Table \ref{table_pw_ign} presents ignition power values for all cells, using a mode from the second pass-band and one from the first dipole band. Measured power values for the second band modes are in exceptional agreement with the one calculated, except for cell 9. This proof of principle experiment for HOMs superposition has given positive results and the HOMs plasma ignition appears to be a promising and feasible method for plasma cleaning of SRF cavities.
When using HOMs to perform in-situ plasma cleaning for LCLS-II cryomodules one should take into account power limitations coming from the cables installed in the cryostat: $\mathrm{P_{FWD}}$ has to be below 20 W  to avoid any possible risks of burning cables. The ignition procedure requires up to 5-10 times the RF power required for plasma cleaning, therefore igniting the plasma once per cavity is preferable to ignite plasma once per cell or nine times per cavity. In addition having a glow discharge always present provides continuous plasma cleaning without interruptions from beginning to end. These are the two main reasons in favor of the plasma bridging for selective cell ignition. The procedure is described in table \ref{plasma_bridge_modes} and has been tested on several 9-cell cavities equipped with LCLS-II HOM couplers. This plasma transfer technique relies on the broad field distribution of certain HOMs, which allows to create an electric field link between cells. This technique has been used at different pressure values for both Ar and Ne, and on different cavities, figure \ref{Ne_cell_1_9} shows a Ne plasma discharge glowing from cell 1 to 9. Regardless of which procedure has been used to ignite the plasma (HOMs superposition or plasma bridging) for a given cell only one mode has been used to sustain the glow discharge: from the picture it is impossible to tell which ignition procedure has been used.

\section{Plasma Tuning and Cell Detection}

\subsection{Plasma tuning and density calculation}
Besides studying the ignition of plasma with different gases and pressures and how to transfer it from cell to cell, it has also been studied how to tune its intensity. The possibility of modifying plasma intensity has been experimentally verified for each pressure. This has been checked both observing the variation of the luminous intensity of the plasma with cameras, but also with the corresponding frequency shift on the VNA. A variation in the plasma intensity corresponds to a change in the dielectric constant of the medium that causes a shift in the resonance frequency of the cavity. The intensity can be tuned varying the forwarded power or the driving frequency $\omega_{rf}$ sent to the cavity.
Using Slater's theorem\cite{Slater} and according to SNS\cite{Doleans_2013plasma} experience it is possible to relate the frequency shift to the intensity of the plasma, intended as the number density (number of particles per cubic meter) of free electrons:
\begin{equation}
    \frac{\delta \omega}{\omega} \approx \frac{1}{2} \frac{\int_{plasma}\eta\,E^2 dV}{\int_{cavity} E^2 dV}
    \label{slater_eq1}
\end{equation}
with
\begin{equation}
 \eta= \frac{\omega^2_{pl}}{\omega^2_{rf}} \, . 
\end{equation}

\begin{figure}
	\centering
	\includegraphics[scale=0.05]{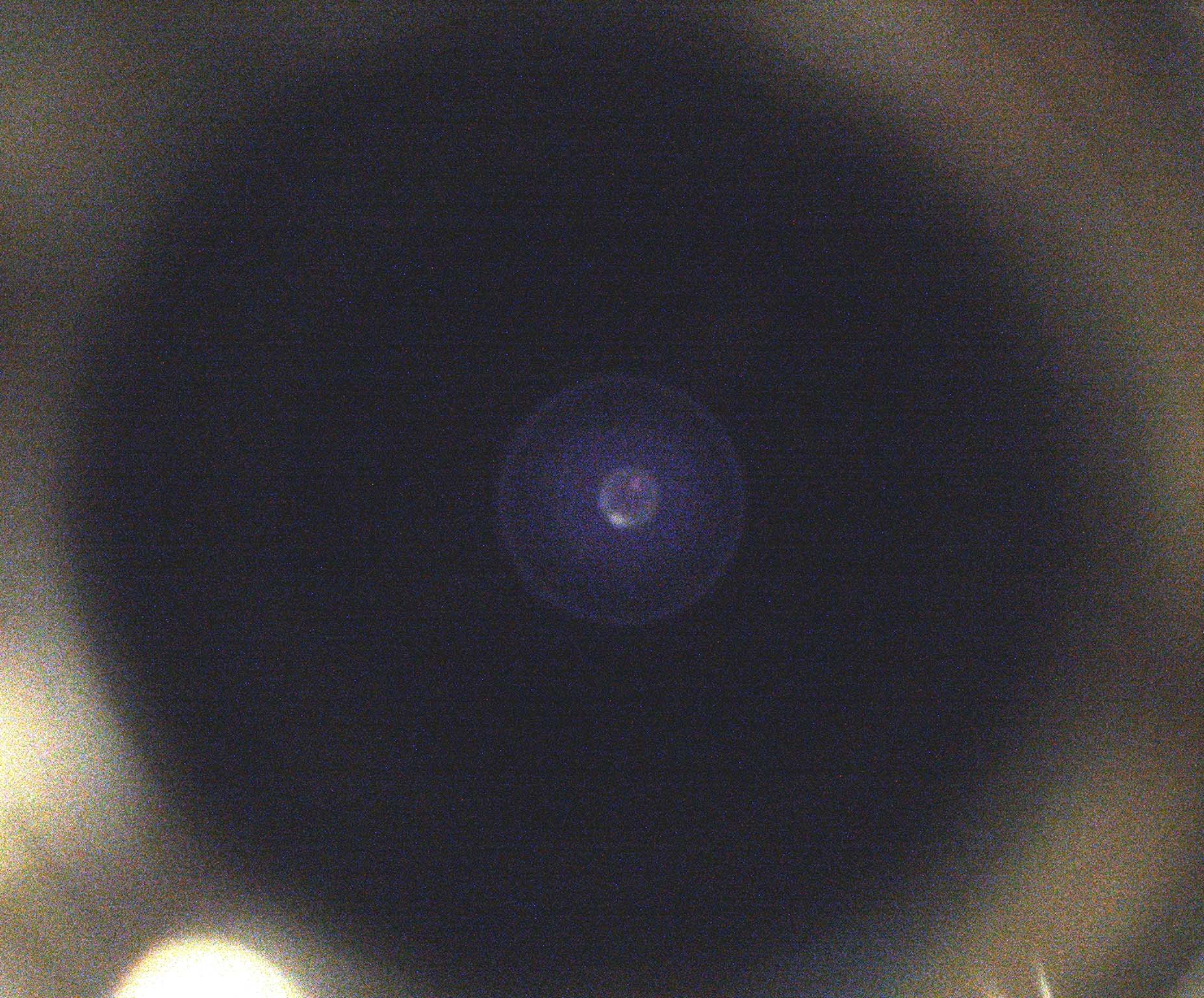}
	\includegraphics[scale=0.05]{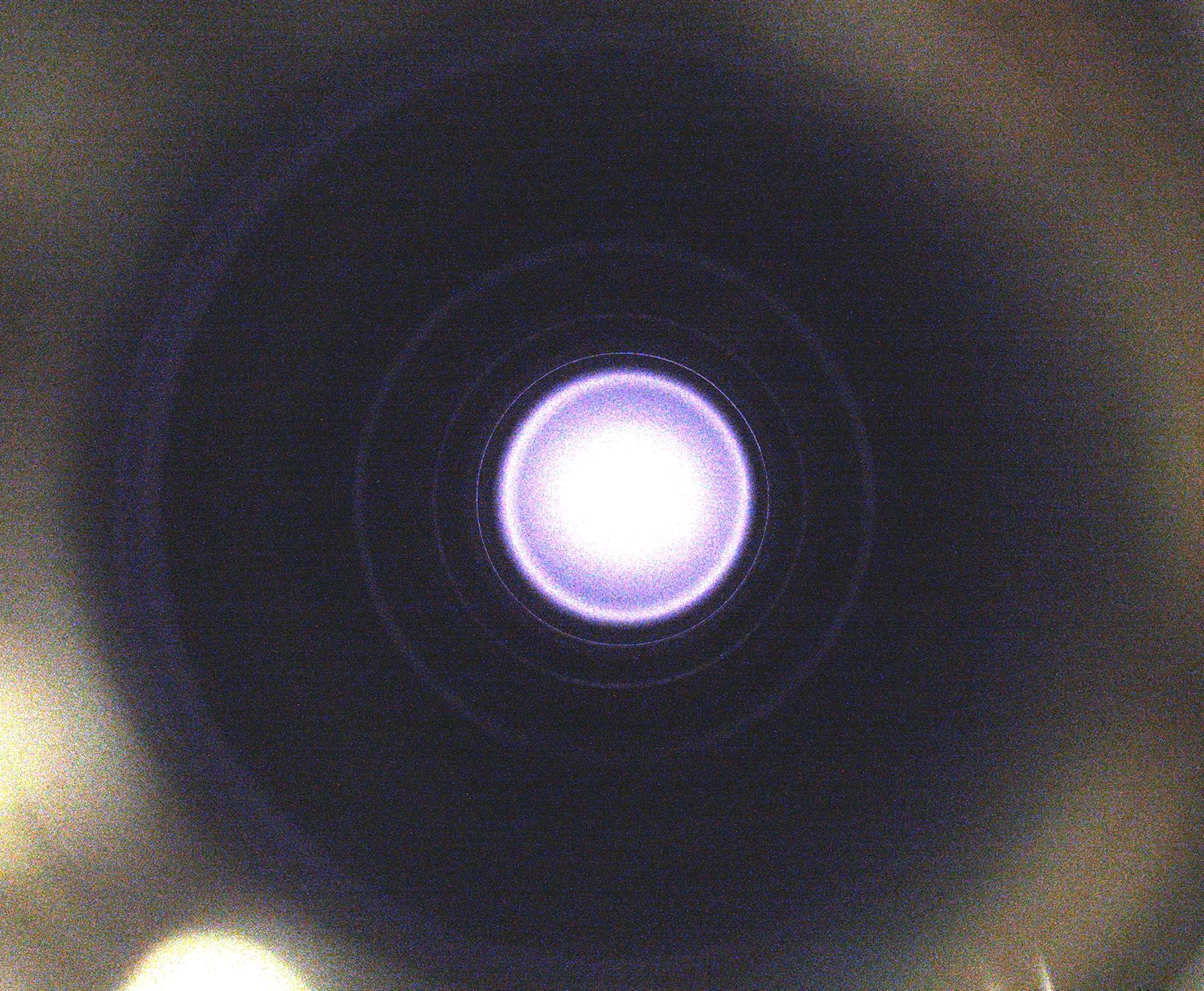}
	\caption{From extremely low (left) to extremely high (right) plasma intensity in the $5^{th}$ cell at 250 mTorr of Argon, mode 1-5.} 
	\label{fig:tuning}

\includegraphics[scale=0.4]{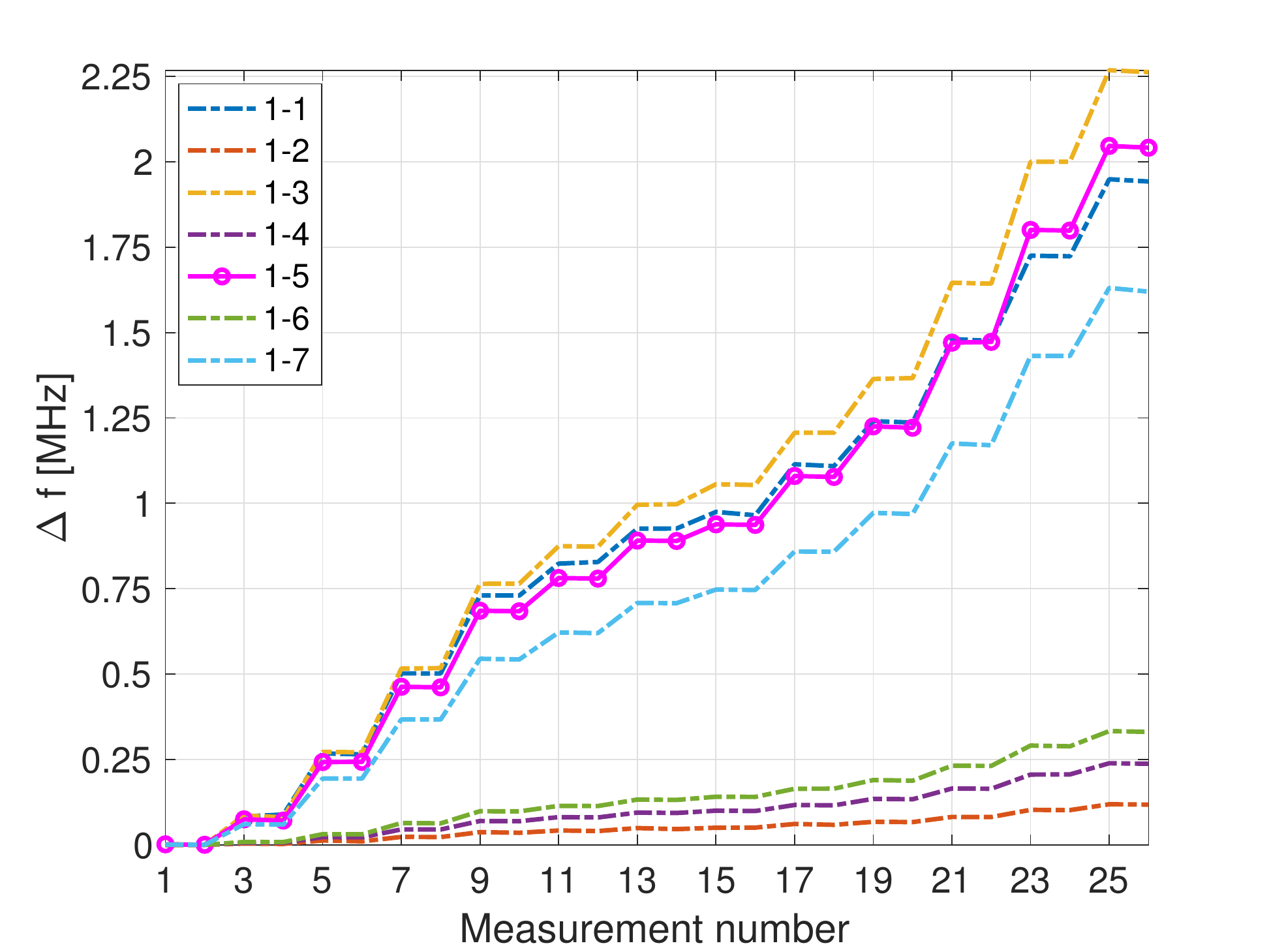}
	\caption{RF frequency measurements of the first dipole pass-band during plasma tuning. The solid line identifies the mode used to sustain the plasma discharge.}
	\label{fig:plot_tuning}
\end{figure}

Using the plasma frequency $\omega_{pl}$ it is possible express the number density $n_e$ in terms of the frequency shift:
\begin{align}
    \omega_{pl} &=\sqrt{\frac{n_e \, e^2}{\epsilon_0 \, m}} \\
    n_e &\approx \frac{\delta \omega}{\omega}\frac{2\varepsilon_0 \, m \omega^2_{rf}}{e^2} \frac{\int_{cavity} E^2 dV}{\int_{plasma} E^2 dV}
    \label{densitynumber}
\end{align}
Once the number density has been estimated it is possible to calculate the number of ionized atoms in the gas sustaining the plasma discharge. \\
The lowest and the highest intensity sustainable by the plasma has been measured. The lowest intensity is identified as the intensity below which the plasma turns off. The maximum instead corresponds to a frequency shift of several MHz (15-20\,MHz), at which the pass-band spectrum is deeply perturbed by the presence of plasma and it is no longer possible to distinguish and excite one resonant mode at the time. Fig.~\ref{fig:tuning} shows the lowest plasma intensity and an example of high plasma intensity at 250\,mTorr of Argon; figure~\ref{fig:plot_tuning} shows the corresponding plot of frequency shift.
Figure \ref{fig:plot_tuning} shows the plot of frequency shift of the dipole pass-band during plasma intensity tuning. The solid curve identifies the frequency shift of mode 1-5, used to ignite the discharge. Each step in frequency shift corresponds to an increase in $\mathrm{P_{FWD}}$. The dashed curves represent the frequency shifts registered by the other modes of the 1st dipole pass-band: modes 1-1, 1-3, 1-7 experience an intense frequency shift since their electric field in the cell of ignition is intense; modes 1-2, 1-4, 1-6 instead show little frequency shift corresponding to low electric field in the ignited cell. Measurements number 1 and 2 correspond to the resonant frequency when plasma is not ignited, 3 and 4 correspond to the lowest plasma intensity shown in Fig.~\ref{fig:tuning}, while 26 and 27 show the maximum plasma intensity reached during this measurement (also shown in Fig.~\ref{fig:tuning}).

The possibility of tuning the intensity of the plasma will allow, during subsequent tests of plasma cleaning, to determine its optimal value to treat hydrocarbon contamination.

\subsection{Detection method and measurements}
\begin{figure}
	\centering
	\includegraphics[scale=0.3]{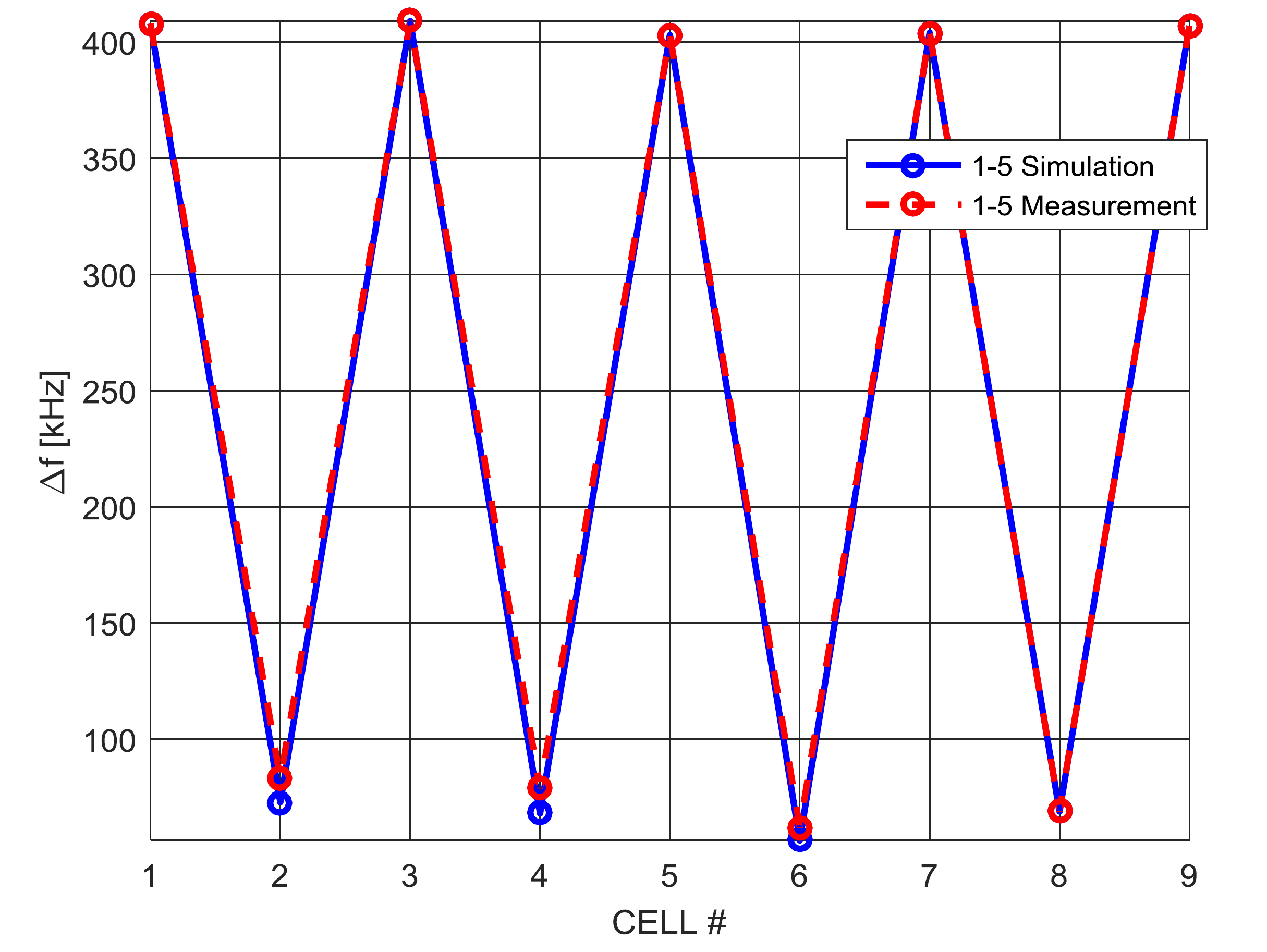}
	\caption{Frequency shift comparison, simulated and measured, for mode 1-5 as a function of plasma location for all cells.} 
	\label{shift_1-5}
\end{figure}

To successfully clean SRF cavities in-situ it is necessary to locate in which cell the plasma glow discharge is located, without having the possibility of looking at the camera image. For SNS HB cavities the detection is done calibrating the transmitted power measurements of the $\mathrm{\pi}$-mode\cite{Dol_RF}. Here a different method is proposed, plasma detection consists in a set of frequency measurements: based on the plasma location the perturbation to different modes is going to induce different frequency shifts. The idea relies on Slater's equation \ref{slater_eq1}, which relates the change in stored electromagnetic energy to the frequency shift for any resonant mode. Using finite elements simulations one can introduce a perturbation in each cavity cell and calculate the frequency shift for all HOMs in presence of plasma, setting the real part of the dielectric constant to less than the unity. Modes from the first dipole pass-band have been chosen to track the plasma location throughout the cavity since measured frequency shifts were more predictable and similar to simulated values. In general it is possible to normalize the frequency shifts since the dielectric constant of the plasma generated during the experiment is unknown. This method has been tested on several 9-cell cavities, the cameras were still used to visually verify the plasma location. Using a volume integral quantity as the stored energy gives the method robustness, even if the HOMs are not tuned like the $\mathrm{\pi}$-mode their stored energy for each cell does not differ significantly from simulated values. Figure \ref{shift_1-5} shows the comparison between calculated and measured frequency shift of mode 1-5 for all cells. Plasma location is identified automatically by a Labview program after measuring the modes from 1-1 to 1-7, and comparing their normalized frequency shifts with calculated values.

\section{Conclusions}
Plasma ignition, tuning, and detection have been demonstrated using HOMs of 9-cell 1.3 GHz cavities. The usage of dipole modes from the $1^{st}$ and $2^{nd}$ pass-band allows overcoming limitations imposed by the coupling of the fundamental monopole band: at room temperature the mismatch between coupler $Q_{ext}\approx 10^7$ and cavity $Q_{0}=10^4$ would require an enormous amount of power to ignite the plasma, imposing the risk of damaging and deteriorating the high power antenna and its parts. LCLS-II cryomodules will be processed in-situ, using this new plasma ignition and control method, which requires only few Watts of RF power flowing through the HOMs cables. Two methods for plasma ignition have been developed: both HOMs superposition and plasma bridging allow dependably moving the glow discharge in each of the cavity cells. Despite the usage of only few Watts of forward power, the plasma remains highly tunable thanks to the very good coupling of HOMs to the cavity also at room temperature. Plasma detection is fundamental for in-situ application of plasma processing. A technique that locates the plasma using simple RF frequency measurements has been developed and has been automated using Labview. All RF aspects of the plasma cleaning for LCLS-II and the technique is ready to be deployed for cryomodule processing. Future work will be highly focused on confirming the effectiveness of plasma cleaning in removing hydrocarbons.

\end{document}